\documentclass[a4paper,fleqn]{cas-sc}

\usepackage[numbers]{natbib}
\usepackage{siunitx}
\usepackage{graphicx}
\usepackage{booktabs}
\usepackage{amsmath}
\usepackage{amssymb}
\usepackage{array}
\usepackage{float}
\usepackage{microtype}
\usepackage{url}
\DisableLigatures{encoding=*,family=*}
\usepackage{tikz}
\usetikzlibrary{arrows.meta,positioning,shapes.geometric,calc}

\ExplSyntaxOn
\cs_set:Npn \__first_footerline:
  {
    \group_begin:
    \small\sffamily
    \ifnum\theblind>0\relax
    \else \__short_authors:
    \fi
    \group_end:
  }
\ExplSyntaxOff

\begin{document}
\let\WriteBookmarks\relax
\def\floatpagepagefraction{1}
\def\textpagefraction{.001}

\shorttitle{Screening bolt loosening with global FRF and local FRAC maps}
\shortauthors{B. Kullukcu et~al.}

\title[mode=title]{Screening bolt loosening in a four-bolt plate with global FRF correlation and local FRAC maps from full-field laser Doppler vibrometry}
\tnotemark[1]

\tnotetext[1]{This work was funded by the European Regional Development Fund (ERDF/EFRE) and the State of Brandenburg within the StaF-Verbund programme, under the project ``Systementwicklung für intelligente und automatisierte mobile Inspektion von Schienenfahrzeugen (SIAMIS)'', administered by the Investment Bank of the State of Brandenburg (ILB), and was supported by the Joint project 6G-life, under Grant 16KISK001K and as part of Germany's Excellence Strategy - EXC 2050/1 - Project ID 390696704 - Cluster of Excellence ``Centre for Tactile Internet with Human-in-the-Loop'' (CeTI) of TU Dresden. The funders had no role in study design, data collection, analysis, interpretation, or manuscript preparation.}
\tnotetext[2]{Abbreviations: LDV, laser Doppler vibrometry; FRF, frequency response function; MAC, modal assurance criterion; CMAC, complex modal assurance criterion; FRAC, frequency response assurance criterion; SHM, structural health monitoring.}

\author[1,2]{Berkay Kullukcu}[orcid=0000-0002-0784-1537]
\cormark[1]
\ead{berkay.kullukcu@th-wildau.de}
\ead[url]{https://www.th-wildau.de}
\credit{Conceptualization, Methodology, Software, Formal analysis, Writing - Original draft}

\author[1]{Robin Pianowski}[orcid=0009-0000-4484-0843]
\ead{robin.pianowski@th-wildau.de}
\credit{Methodology, Formal analysis, Writing - Original draft}

\author[2]{Mehmet Sait Özer}[orcid=0000-0001-7244-3503]
\ead{mehmet_sait.oezer@tu-dresden.de}
\credit{Formal analysis, Writing - Original draft}

\author[2,3,4]{Ercan Altinsoy}[orcid=0000-0002-0803-8818]
\ead{ercan.altinsoy@tu-dresden.de}
\credit{Supervision, Resources}

\author[1]{Dina Hannebauer}[orcid=0009-0000-1572-2215]
\ead{dina.hannebauer@th-wildau.de}
\credit{Supervision, Resources, Writing - Review \& editing}

\cortext[cor1]{Corresponding author}

\affiliation[1]{organization={FG Machine Dynamics and Acoustics, TH Wildau},
                addressline={Hochschulring 1},
                city={Wildau},
                postcode={15745},
                state={Brandenburg},
                country={Germany}}

\affiliation[2]{organization={Institute of Acoustics and Speech Communication, Chair of Acoustics and Haptics, TU Dresden},
                postcode={01069},
                state={Saxony},
                country={Germany}}

\affiliation[3]{organization={Centre for Tactile Internet with Human-in-the-Loop (CeTI), TU Dresden},
                postcode={01069},
                state={Saxony},
                country={Germany}}

\affiliation[4]{organization={Research Cluster 6G-life, TU Dresden},
                postcode={01069},
                state={Saxony},
                country={Germany}}

\begin{abstract}
Full-field laser Doppler vibrometry (LDV) can reveal how bolt torque loss redistributes a frequency response function (FRF) over an entire structure rather than only at a few sensor positions. This work presents a screening procedure for a four-bolt aluminum plate using pointwise amplitude and phase exports of scanned $H_1$ FRFs. Candidate resonances were selected from the all-tight spatial RMS spectrum and tracked in four single-bolt \SI{0}{\newton\metre} cases. Seven baseline responses remained trackable within prescribed group-specific search windows and were retained as experimental resonance groups RG1$\--$RG7; a dimensionless plate-frequency coefficient was reported alongside each measured frequency. Global changes were quantified using amplitude-only and complex, phase-retaining modal-assurance dissimilarities, and local changes were evaluated using matched-window FRAC-deficit maps and a normalized hotspot-area fraction. The retained groups separate into low-, intermediate-, and high-distortion responses, while the local maps distinguish compact joint-centered changes from distributed FRF redistribution. The workflow therefore links global FRF distortion with spatially resolved interpretation without relying on a trained classifier or specimen-specific node labels.
\end{abstract}


\maketitle

\section{Introduction}
Bolted joints influence structural dynamics through several coupled mechanisms. Frictional slip at the contact interfaces can soften resonance frequencies and increase damping \cite{Wall2022}. Preload-dependent contact nonlinearity can also modify the higher-order frequency-response behavior of an assembled structure \cite{Teloli2021}, while joint degradation and bolt-assembly conditions can alter effective stiffness and the measured vibration response \cite{Shi2024}. A recent review summarizes the broader range of threaded-fastener loosening-detection strategies \cite{Huang2022Review}. Representative approaches include contact-acoustic-nonlinearity features based on higher harmonics and spectral sidebands \cite{Zhang2018CAN}, time-reversal vibro-acoustic modulation \cite{WangSong2019VAM}, guided-wave phase indicators \cite{Tong2025PhaseLamb}, and feature-based machine-learning classification for multi-bolt connections \cite{Wang2020ML}.

These methods provide damage-sensitive variables, but they do not necessarily provide a directly measured, dense map of how the structural response changes over the inspected surface. Continuous-scanning laser Doppler vibrometry provides non-contact vibration measurements with high spatial sampling density \cite{DiMaio2021}. Full-field optical measurements have been used for linear and nonlinear dynamic characterization \cite{Ehrhardt2017} and for plate-damage identification from spatial vibration fields \cite{Chen2018PlateLDV}. The resulting response field supports two complementary levels of comparison. The modal assurance criterion (MAC) quantifies similarity between spatial response vectors \cite{AllemangBrown1982,Gres2021}. Complex-valued assurance and frequency-domain correlation formulations retain phase information that is discarded by an amplitude-only comparison \cite{Vacher2010,Perez2019}. At the local level, the frequency response assurance criterion (FRAC) compares two FRFs over a selected frequency range \cite{ChenFRAC2003}. Spatial FRF-shape methods can support damage-sensitive localization \cite{Liu2009}, while recent complex frequency-domain correlation approaches have been applied to damage quantification from measured FRFs \cite{Perez2021Ice}.

\begin{figure}[t]
\centering
\resizebox{0.96\textwidth}{!}{%
\begin{tikzpicture}[
    font=\footnotesize,
    >=Latex,
    node distance=6.5mm and 10mm,
    data/.style={trapezium, trapezium left angle=70, trapezium right angle=110,
                 draw=black, fill=gray!10, align=center, text width=34mm,
                 minimum height=8mm, inner sep=2pt},
    process/.style={rounded corners=2pt, draw=black, fill=gray!18,
                    align=center, text width=36mm, minimum height=8mm, inner sep=2.5pt},
    metric/.style={rounded corners=2pt, draw=black, fill=gray!30,
                   align=center, text width=38mm, minimum height=9mm, inner sep=2.5pt},
    decision/.style={diamond, aspect=2.1, draw=black, fill=gray!8,
                     align=center, text width=29mm, inner sep=1.5pt},
    output/.style={rounded corners=2pt, draw=black, fill=gray!42,
                   align=center, text width=52mm, minimum height=9mm, inner sep=2.5pt},
    arrow/.style={->, line width=0.6pt},
    reject/.style={->, line width=0.5pt, dashed}
]

\node[data] (exports) {LDV $H_1$ FRF exports\\amplitude $|H_i(f)|$ and phase $\phi_i(f)$};
\node[process, below=of exports] (common) {Coordinate harmonization\\51 common LDV points};
\node[process, below=of common] (srms) {All-tight spatial RMS spectrum\\$S(f)=\sqrt{N^{-1}\sum_i |H_i(f)|^2}$};
\node[process, below=of srms] (candidates) {Peak picking and peak-width de-duplication\\candidate resonance groups};
\node[decision, below=of candidates] (tracking) {Peak found within the\\prescribed search window in\\all single-bolt \SI{0}{\newton\metre} cases?};
\node[process, below=of tracking] (retained) {Retained set\\RG1--RG7};

\node[metric, below left=10mm and 20mm of retained] (global) {Global response change\\$1-\mathrm{MAC}_a$ and $1-\mathrm{CMAC}$};
\node[metric, below right=10mm and 20mm of retained] (local) {Local response change\\FRAC-deficit maps};

\node[process, below=of global] (globalrank) {Global distortion ranking};
\node[process, below=of local] (hotspot) {Normalized hotspot area\\$\eta_{0.75}$};

\coordinate (screenmerge) at ($(globalrank.south)!0.5!(hotspot.south)$);
\node[output, below=10mm of screenmerge] (interpretation) {Screening interpretation\\compact joint-centered change versus distributed FRF redistribution};
\coordinate (leftinput) at ($(interpretation.north)+(-13mm,0)$);
\coordinate (rightinput) at ($(interpretation.north)+(13mm,0)$);

\draw[arrow] (exports) -- (common);
\draw[arrow] (common) -- (srms);
\draw[arrow] (srms) -- (candidates);
\draw[arrow] (candidates) -- (tracking);
\draw[arrow] (tracking) -- node[right, xshift=1mm] {yes} (retained);
\draw[reject] (tracking.east) -- ++(20mm,0) node[right, align=left] {no: excluded\\from the retained set};
\draw[arrow] (retained.south) -- ++(0,-5mm) -| (global.north);
\draw[arrow] (retained.south) -- ++(0,-5mm) -| (local.north);
\draw[arrow] (global) -- (globalrank);
\draw[arrow] (local) -- (hotspot);
\draw[arrow] (globalrank.south) -- ++(0,-5mm) -| (leftinput);
\draw[arrow] (hotspot.south) -- ++(0,-5mm) -| (rightinput);

\end{tikzpicture}%
}
\caption{Focused LDV-based screening workflow. Resonance groups are selected from the all-tight spatial RMS spectrum, filtered by window-constrained trackability in the four single-bolt \SI{0}{\newton\metre} cases, ranked by global FRF dissimilarity, and interpreted through local FRAC-deficit maps and normalized hotspot area.}
\label{fig:workflow}
\end{figure}
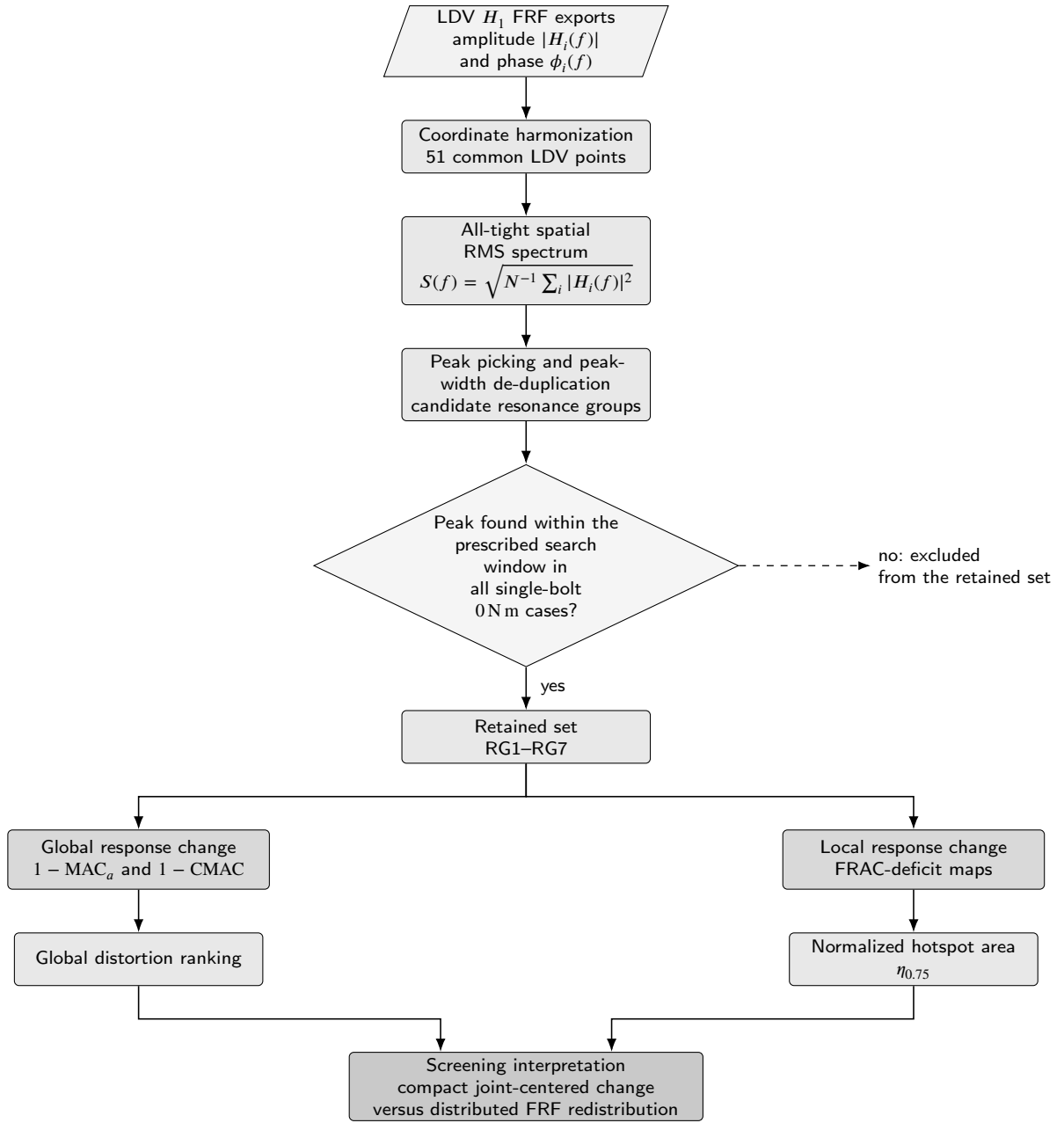

A remaining difficulty is transferability. A raw frequency label is specimen-specific, whereas assigning an FRF peak a mode number requires modal-parameter identification rather than peak picking alone \cite{Reynders2012}. To separate these issues, the present study uses ordered experimental resonance-group labels and reports one dimensionless plate-frequency coefficient alongside each measured frequency. The RG label preserves the identity of a selected all-tight response region, while the normalized coefficient supplies frequency information scaled by plate dimensions and bending stiffness, following established thin-plate vibration practice \cite{Senjanovic2015,Eftekhari2021}. The contribution is a screening workflow that links resonance-group selection, amplitude-only and complex global FRF comparison, and local FRAC-based spatial interpretation. The full procedure is summarized in Fig.~\ref{fig:workflow}.

\section{Experimental setup and analysis workflow}
The analysis comprises three stages. First, baseline resonance groups are selected from the all-tight spatial RMS spectrum and retained when a corresponding peak can be located within the prescribed group-specific search window in all four single-bolt \SI{0}{\newton\metre} cases. Second, amplitude-only and complex, phase-retaining global FRF dissimilarities are used to rank the retained groups. Third, local FRAC-deficit maps and a normalized hotspot-area fraction are used to distinguish compact joint-centered changes from distributed FRF redistribution.
\subsection{Specimen and measurements}
\begin{figure}[t]
\centering
\includegraphics[width=0.96\textwidth]{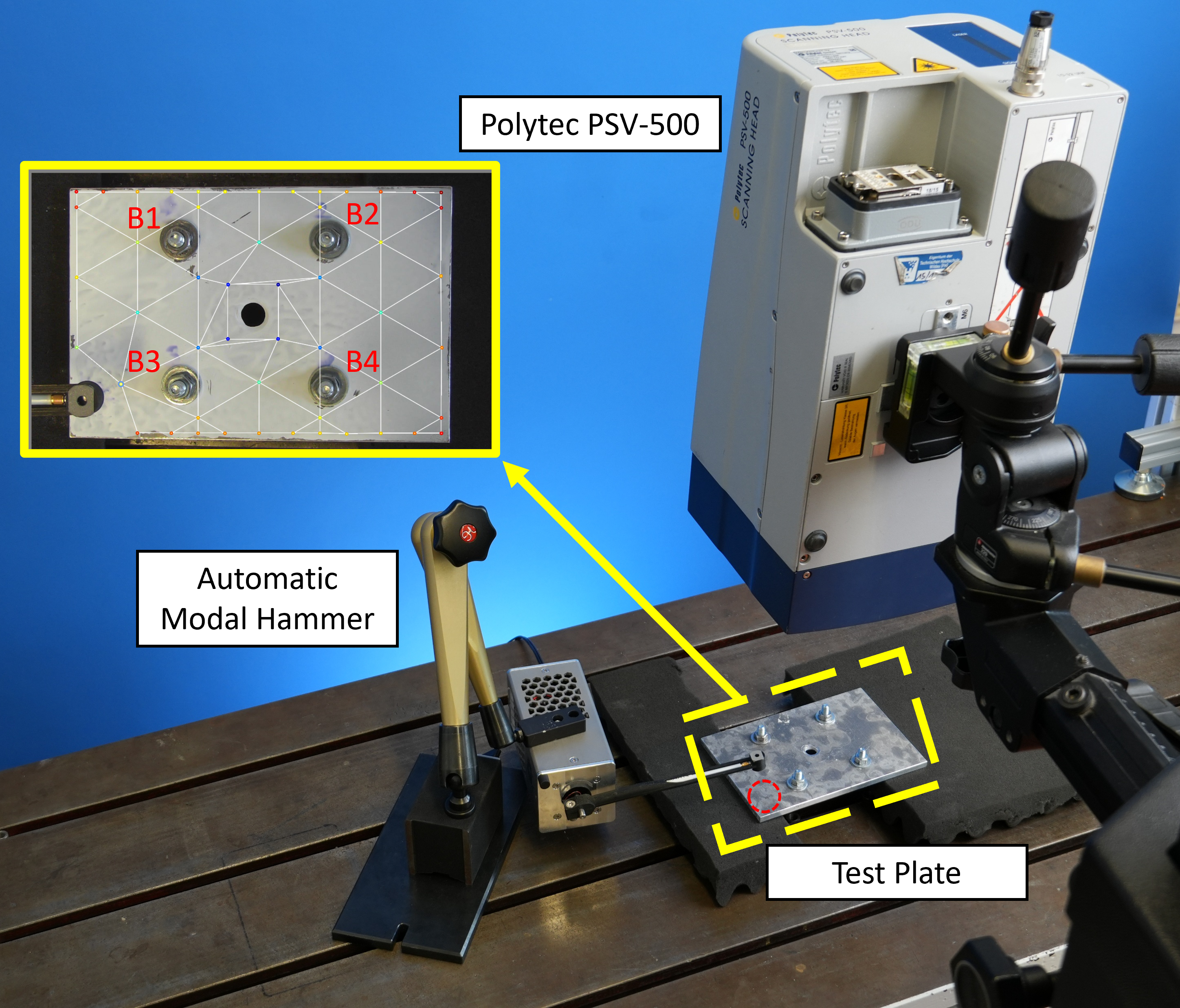}
\caption{Experimental setup for the full-field LDV measurements. The four-bolt aluminum plate was supported on soft
sponges to approximate free-free boundary conditions and excited at the lower-left edge (highlighted on top of the plate
with a red circle), near bolt 3, using the automatic modal hammer. The Polytec PSV-500 scanning laser
Doppler vibrometer measured pointwise $H_1$ displacement-to-force FRFs over the plate surface; the inset indicates the LDV
grid region and bolt labels used throughout the analysis.}
\label{fig:test}
\end{figure}

The specimen is a $100\times150\times7$~mm aluminum plate with four diagonally placed M6 bolts and an unused M6 hole at its center. The bolts are labeled 1$\--$4, with bolts 1 and 4 on one diagonal and bolts 2 and 3 on the other. The complete database contains 18 torque configurations. The screening analysis reported here uses the all-tight reference, in which all bolts were tightened to \SI{10}{\newton\metre}, and four single-bolt \SI{0}{\newton\metre} cases, in which one bolt was loose while the remaining three stayed at \SI{10}{\newton\metre}. The remaining measured configurations are retained in the published database but are outside the scope of the present study. The plate was supported on soft sponges to approximate free-free boundary conditions and excited up to \SI{20}{\kilo\hertz} with an automatic modal hammer. Impacts were applied at the lower-left edge near bolt 3. A Polytec PSV-500 scanning laser Doppler vibrometer measured pointwise $H_1$ displacement-to-force FRFs in amplitude and phase. The setup is shown in Fig.~\ref{fig:test}. Of the 79 measured LDV positions, 51 coordinates were common to the amplitude and phase exports of all analyzed states and were used for the comparisons.

For a given torque state, the amplitude vector at frequency $f$ is
\begin{equation}
\mathbf{a}(f)=
\begin{bmatrix}
|H_1(f)| & |H_2(f)| & \ldots & |H_N(f)|
\end{bmatrix}^{\top},
\label{eq:ampvec}
\end{equation}
where $H_i(f)$ is the pointwise displacement-to-force $H_1$ FRF at LDV point $i$ and $N=51$. The spatial RMS spectrum is
\begin{equation}
S(f)=\sqrt{\frac{1}{N}\sum_{i=1}^{N}|H_i(f)|^2}.
\label{eq:srms}
\end{equation}
The exports span \SI{1}{\hertz}$\--$\SI{20}{\kilo\hertz} with a frequency resolution of \SI{1}{\hertz}. The resulting all-tight spatial RMS spectrum used for resonance-group selection is shown in Fig.~\ref{fig:srmspeaks}. The $H_1$ estimator is a standard spectral FRF estimate for measured input$\--$output vibration data \cite{MaoTodd2013}.

\subsection{Resonance-group selection and tracking}
The retained responses are indexed RG1$\--$RG7 in order of increasing all-tight frequency. The labels are used as experimental response-group identifiers. To provide one geometry- and material-normalized descriptor, each all-tight frequency $f_g$ is also reported through
\begin{equation}
\Omega_{a,g}=2\pi f_g a^2\sqrt{\frac{\rho h}{D}},
\qquad
D=\frac{Eh^3}{12(1-\nu^2)},
\label{eq:dimfreq}
\end{equation}
where $a=0.10$\,m is the short plate dimension, $h=0.007$\,m is the plate thickness, and $D$ is the Kirchhoff$\--$Love bending stiffness. The coefficient follows the dimensionless frequency scaling used in thin-plate vibration analysis \cite{Senjanovic2015,Eftekhari2021}. The alloy grade was not documented in the exported dataset. Representative nominal aluminum properties, $E=69$\,GPa, $\rho=2700$\,kg\,m$^{-3}$, and $\nu=0.33$, were therefore adopted only for calculating $\Omega_a$; these values are analysis assumptions rather than identified specimen properties, and the measured FRF-based metrics do not depend on them.

\begin{figure}[t]
\centering
\includegraphics[width=0.96\textwidth]{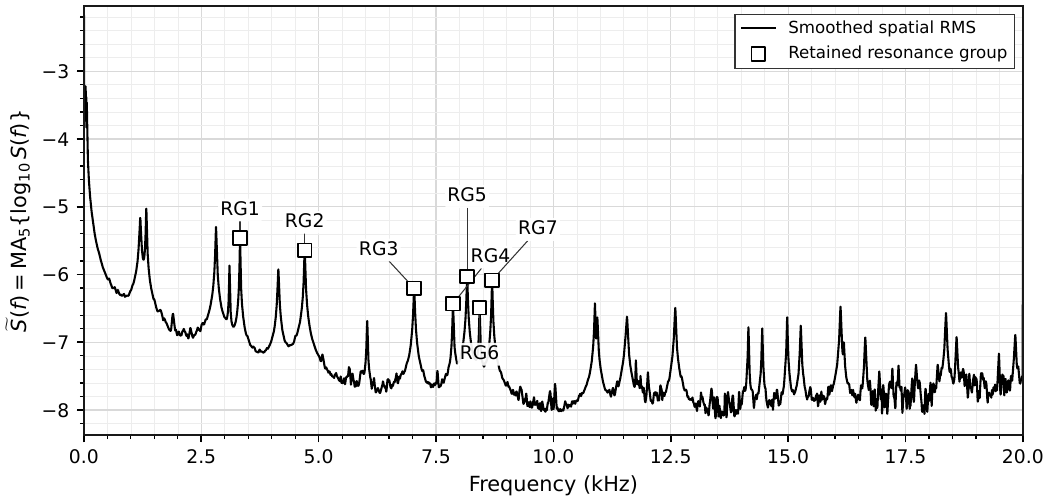}
\caption{All-tight spatial RMS spectrum used for resonance-group selection. The line shows the five-point smoothed logarithmic spatial RMS spectrum $\widetilde{S}(f)$ over \SI{1}{\hertz}$\--$\SI{20}{\kilo\hertz}. Square markers indicate the seven retained resonance groups RG1$\--$RG7; the complete candidate pool is not shown.}
\label{fig:srmspeaks}
\end{figure}

Candidate groups were selected from the all-tight $S(f)$. A five-point moving average was applied to the logarithmic spectrum,
\begin{equation}
\widetilde{S}(f_j)=\frac{1}{5}\sum_{m=-2}^{2}\log_{10}S(f_{j+m}),
\label{eq:smooth}
\end{equation}
and local maxima were characterized by their prominence $P_m$ and peak width at half prominence $B_m$. Neighboring maxima were treated as one broadened group when
\begin{equation}
|f_p-f_q| < \frac{1}{2}(B_p+B_q),
\label{eq:dedup}
\end{equation}
with only the more prominent maximum retained. After this de-duplication, the primary candidate pool comprised the 15 most prominent maxima over the full measurement band. To avoid excluding closely spaced neighboring responses solely because of their global prominence rank, the remaining de-duplicated maxima in the dense 7.7$\--$8.8~kHz band were also admitted before the trackability screening. Figure~\ref{fig:srmspeaks} displays only the seven groups retained for the subsequent global and local analysis; it does not display the complete candidate pool. Seven baseline groups had a comparison-case peak within the prescribed search region in all four single-bolt \SI{0}{\newton\metre} cases and were retained as RG1$\--$RG7. For each group $g$, the corresponding comparison-case frequency was defined as the most prominent local maximum within a symmetric window centered on the all-tight group frequency $f_g$. The implemented tracking half-width was
\begin{equation}
\Delta f_g=
\begin{cases}
120\,\mathrm{Hz}, & g=\mathrm{RG1},\\
180\,\mathrm{Hz}, & g=\mathrm{RG2},\ldots,\mathrm{RG7}.
\end{cases}
\label{eq:trackingwindow}
\end{equation}
These implemented windows accommodate the observed peak relocation while retaining an association with the corresponding all-tight frequency region. In the densely populated high-frequency band, the windows of adjacent baseline groups overlap. Adjacent RGs can therefore select the same comparison-case local maximum. The RG labels identify all-tight baseline response groups and their window-constrained counterparts; they do not imply a one-to-one identification of distinct analytical modes in every loose-bolt case. The tracked frequencies and normalized coefficients are listed in Table~\ref{tab:tracked}.

\begin{table}[t]
\centering
\caption{Retained all-tight resonance groups, short-side plate-normalized frequency coefficient, and window-constrained single-bolt loose-case peak frequencies. In the densely spaced high-frequency band, overlapping windows can assign the same comparison-case local maximum to adjacent baseline groups.}
\label{tab:tracked}
\small
\setlength{\tabcolsep}{3.5pt}
\begin{tabular}{lcccccc}
\toprule
Group & \shortstack{All-tight\\frequency (Hz)} & $\Omega_a$ & \shortstack{Bolt 1\\loose} & \shortstack{Bolt 2\\loose} & \shortstack{Bolt 3\\loose} & \shortstack{Bolt 4\\loose} \\
\midrule
RG1 & 3325 & 19.31 & 3402 & 3404 & 3399 & 3410 \\
RG2 & 4703 & 27.31 & 4837 & 4872 & 4842 & 4868 \\
RG3 & 7034 & 40.84 & 7105 & 7117 & 7119 & 7130 \\
RG4 & 7861 & 45.64 & 8007 & 7989 & 8038 & 7985 \\
RG5 & 8163 & 47.40 & 8238 & 7989 & 8220 & 7985 \\
RG6 & 8428 & 48.93 & 8248 & 8541 & 8564 & 8294 \\
RG7 & 8695 & 50.49 & 8875 & 8541 & 8564 & 8875 \\
\bottomrule
\end{tabular}
\end{table}

\subsection{Global and local screening metrics}
Global response-field similarity was first evaluated with the amplitude-only modal assurance criterion (MAC). At each tracked peak, the all-tight amplitude vector $\mathbf{a}_b$ was compared with the corresponding loose-case vector $\mathbf{a}_c$ using
\begin{equation}
\mathrm{MAC}_a(\mathbf{a}_b,\mathbf{a}_c)=
\frac{|\mathbf{a}_b^{\top}\mathbf{a}_c|^2}
{(\mathbf{a}_b^{\top}\mathbf{a}_b)(\mathbf{a}_c^{\top}\mathbf{a}_c)},
\label{eq:maca}
\end{equation}
reported as the dissimilarity $1-\mathrm{MAC}_a$ \cite{AllemangBrown1982,Gres2021}. The complex FRF vector is
\begin{equation}
\mathbf{z}(f)=\mathbf{a}(f)\odot e^{\mathrm{i}\boldsymbol{\phi}(f)},
\label{eq:zvec}
\end{equation}
where $\boldsymbol{\phi}$ contains the measured phase angles. The complex modal assurance criterion (CMAC), which uses complex FRF vectors and therefore retains the measured phase information, is
\begin{equation}
\mathrm{CMAC}(\mathbf{z}_b,\mathbf{z}_c)=
\frac{|\mathbf{z}_b^{\mathrm{H}}\mathbf{z}_c|^2}
{(\mathbf{z}_b^{\mathrm{H}}\mathbf{z}_b)(\mathbf{z}_c^{\mathrm{H}}\mathbf{z}_c)},
\label{eq:cmac}
\end{equation}
reported as the complex-valued dissimilarity $1-\mathrm{CMAC}$ \cite{Vacher2010,Perez2019}. Group-wise means and 95\% exhaustive-bootstrap intervals were calculated over the four single-bolt cases. All $4^4=256$ ordered resamples of size four with replacement were evaluated. The arithmetic mean was calculated for each resample, and the 2.5th and 97.5th percentiles of the resulting mean distribution define the reported interval.

For local comparison, equal-width frequency windows were extracted around the all-tight and corresponding loose-case tracked peaks. At point $i$, the local FRAC was
\begin{equation}
\mathrm{FRAC}_i=
\frac{\left|\sum_k H_{b,i}^{*}(f_{b,k})H_{c,i}(f_{c,k})\right|^2}
{\left(\sum_k|H_{b,i}(f_{b,k})|^2\right)
 \left(\sum_k|H_{c,i}(f_{c,k})|^2\right)},
\label{eq:localfrac}
\end{equation}
with local deficit
\begin{equation}
L_i=1-\mathrm{FRAC}_i.
\label{eq:fracdeficit}
\end{equation}
A hotspot is a spatial region in which the matched-window loose-case FRF departs strongly from the all-tight FRF. The FRAC formulation follows frequency-domain response-correlation practice \cite{ChenFRAC2003}. Related complex frequency-domain and FRF-shape approaches support damage-sensitive localization \cite{Perez2019,Liu2009,Perez2021Ice}. To quantify hotspot extent, the common LDV coordinates were projected onto their best-fit plate plane and their convex hull defined the scanned area $A_{\mathrm{scan}}=98.8$~cm$^2$ \cite{Barber1996}. For each single-bolt case, the projected deficit values were reflected so that the loosened bolt occupied a common reference quadrant and were interpolated using piecewise-linear scattered-data interpolation on a $500\times350$ rectangular grid. Grid points outside the convex hull, or outside the common valid interpolation region of the four aligned cases, were excluded. The four aligned grids were then averaged pointwise. For group $g$, the mean deficit field was min-max normalized,
\begin{equation}
\widehat{L}_g(\mathbf{r})=
\frac{\overline{L}_g(\mathbf{r})-\min_{A_{\mathrm{scan}}}\overline{L}_g}
{\max_{A_{\mathrm{scan}}}\overline{L}_g-\min_{A_{\mathrm{scan}}}\overline{L}_g},
\label{eq:normhotspot}
\end{equation}
and the normalized high-deficit area was
\begin{equation}
\eta_{g,\tau}=\frac{1}{A_{\mathrm{scan}}}
\int_{A_{\mathrm{scan}}}\mathbb{1}\!\left[\widehat{L}_g(\mathbf{r})\ge\tau\right] \,\mathrm{d}A,
\qquad \tau=0.75.
\label{eq:hotspotarea}
\end{equation}
The integral in Eq.~\eqref{eq:hotspotarea} was evaluated by counting the valid in-hull grid cells that satisfied the threshold and multiplying by the grid-cell area. The fixed value $\tau=0.75$ was used as a descriptive high-deficit-core threshold; it was not optimized from the present dataset, and no threshold-sensitivity analysis was used to select it.

\section{Results}
Fig.~\ref{fig:srmspeaks} shows the all-tight spatial RMS spectrum and the seven retained resonance groups. Their all-tight frequencies span 3325$\--$8695~Hz, corresponding to $\Omega_a=19.31$ $\--$50.49. For group $g$ and loose-bolt case $c$, the tracked-frequency change is defined as $\Delta f_{g,c}=f_{g,c}-f_g$. Table~\ref{tab:tracked} shows that the lower-frequency groups exhibit consistent window-constrained peak relocation after bolt loosening. For RG1$\--$RG4, all four single-bolt cases move to higher tracked frequencies, with $\Delta f=74$ $\--$85~Hz for RG1, 134$\--$169~Hz for RG2, 71$\--$96~Hz for RG3, and 124$\--$177~Hz for RG4. The behavior becomes less uniform in the dense band above \SI{8}{\kilo\hertz}. RG5$\--$RG7 exhibit both upward and downward relocation depending on which bolt is loose, with case-wise changes extending from approximately $\Delta f=-180$ to $+180$~Hz. This bidirectional relocation is consistent with the close spacing of neighboring response groups in this band and motivates interpreting the retained responses as tracked experimental resonance groups rather than assigning analytical mode numbers from peak frequency alone.

\begin{figure}[t]
\centering
\includegraphics[width=0.98\textwidth]{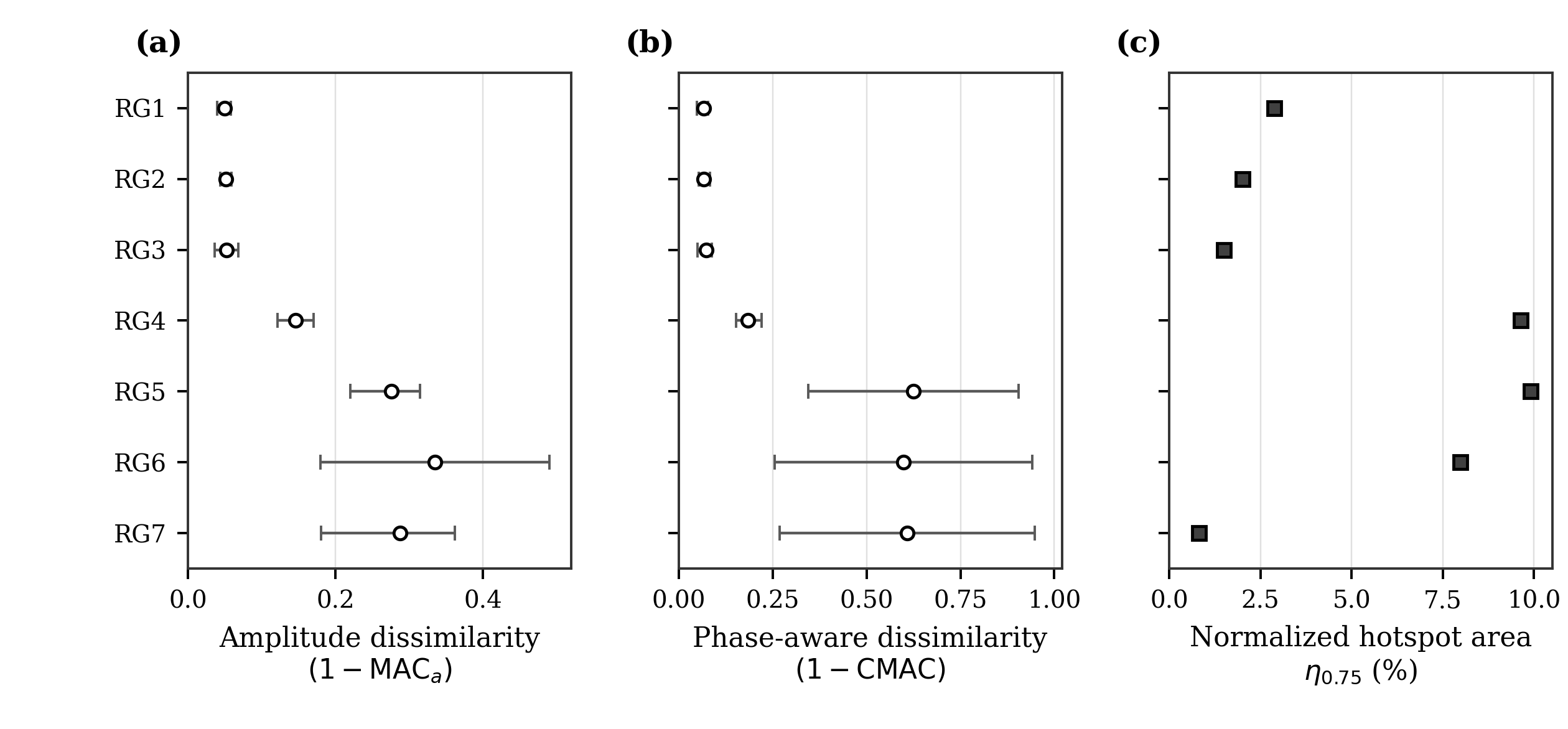}
\caption{Focused screening metrics for RG1$\--$RG7. Panels (a) and (b) show mean amplitude-only and phase-aware global dissimilarity; horizontal bars denote 95\% exhaustive-bootstrap intervals. Panel (c) shows the normalized area of the $\widehat{L}_g\ge0.75$ hotspot core.}
\label{fig:metrics}
\end{figure}

The global screening results are summarized in Table~\ref{tab:summary} and Fig.~\ref{fig:metrics}(a)$\--$(b). RG1$\--$RG3 form a closely grouped low-distortion set. Their mean amplitude dissimilarities, $1-\mathrm{MAC}_a$, lie between 0.049 and 0.052, while their phase-aware dissimilarities, $1-\mathrm{CMAC}$, lie between 0.066 and 0.073. RG4 separates from this set, with $1-\mathrm{MAC}_a=0.146$ and $1-\mathrm{CMAC}=0.185$, and therefore occupies an intermediate position. The largest response-field changes occur for RG5$\--$RG7. Their amplitude dissimilarities are 0.276, 0.335, and 0.288, respectively, whereas the corresponding phase-aware dissimilarities reach 0.624, 0.599, and 0.608. The increase from the amplitude-only to the complex-valued measure is therefore substantially larger for RG5$\--$RG7 than for RG1$\--$RG4. Because each comparison is performed at the corresponding tracked peak rather than at one fixed absolute frequency, this separation cannot be attributed only to resonance relocation; it indicates a pronounced redistribution of the complex spatial FRF pattern. The uncertainty intervals also broaden for the high-frequency groups, particularly for $1-\mathrm{CMAC}$, showing that the magnitude of this redistribution depends strongly on which bolt is loosened.

\begin{table}[t]
\centering
\caption{Global dissimilarities and normalized hotspot-area fraction for the seven retained resonance groups. Intervals are 95\% exhaustive-bootstrap intervals over the four single-bolt \SI{0}{\newton\metre} cases.}
\label{tab:summary}
\small
\setlength{\tabcolsep}{6pt}
\begin{tabular}{lccc}
\toprule
Group & $1-\mathrm{MAC}_a$ & $1-\mathrm{CMAC}$ & $\eta_{0.75}$ (\% of scanned area) \\
\midrule
RG1 & 0.049 (0.040, 0.058) & 0.066 (0.049, 0.079) & 2.87 \\
RG2 & 0.051 (0.043, 0.059) & 0.068 (0.054, 0.083) & 2.01 \\
RG3 & 0.052 (0.036, 0.068) & 0.073 (0.050, 0.088) & 1.49 \\
RG4 & 0.146 (0.121, 0.170) & 0.185 (0.153, 0.220) & 9.63 \\
RG5 & 0.276 (0.220, 0.314) & 0.624 (0.344, 0.904) & 9.91 \\
RG6 & 0.335 (0.180, 0.490) & 0.599 (0.256, 0.942) & 7.98 \\
RG7 & 0.288 (0.181, 0.362) & 0.608 (0.268, 0.948) & 0.81 \\
\bottomrule
\end{tabular}
\end{table}

\begin{figure}[t]
\centering
\includegraphics[width=0.82\textwidth]{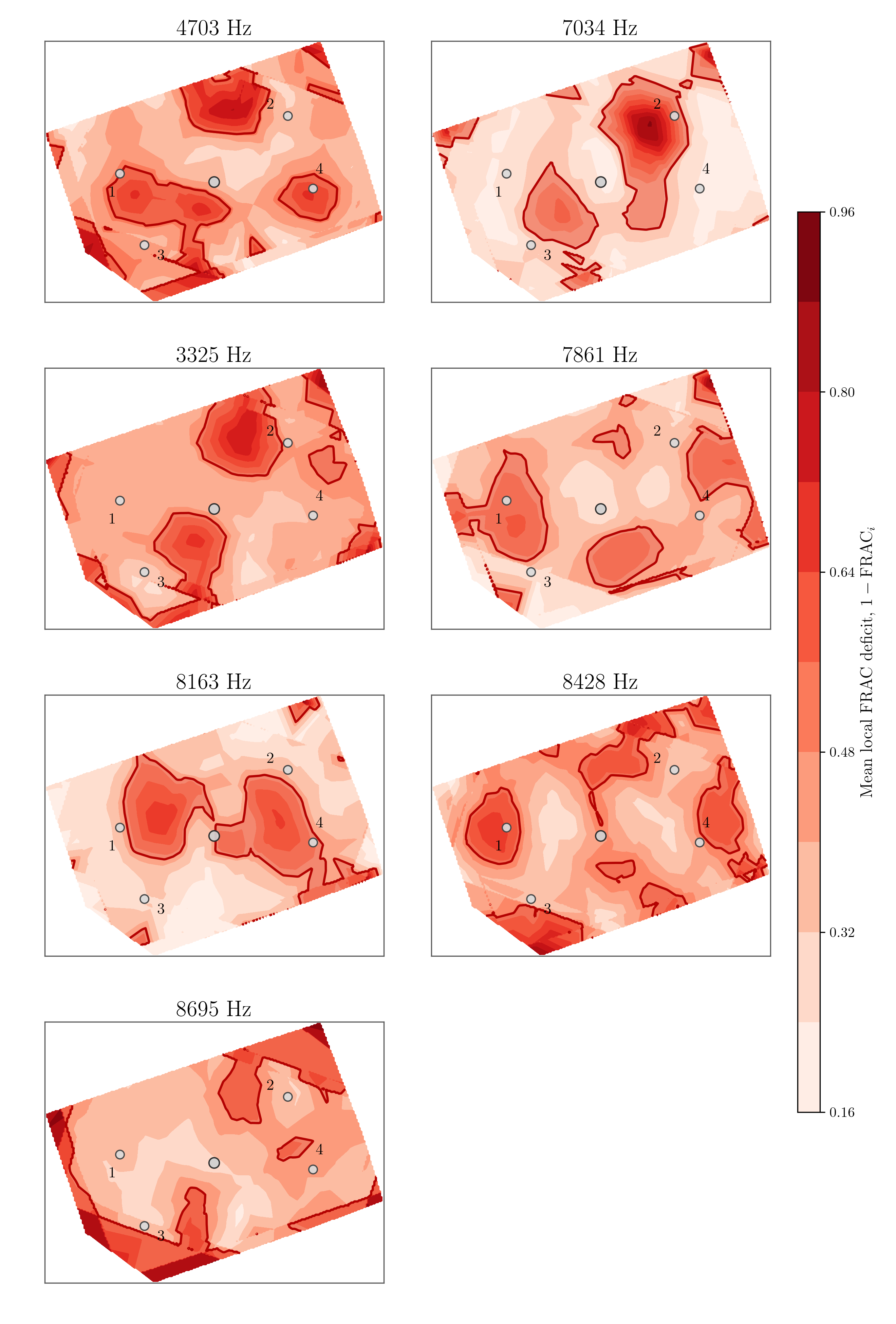}
\caption{Joint-centered mean local FRAC-deficit fields arranged sequentially from RG1 to RG7, with the corresponding all-tight frequency shown in each panel title. The four single-bolt maps were reflected to align the loosened-bolt quadrant and then averaged. Numbered circles indicate bolt positions after alignment. Darker regions denote larger local FRAC deficit, and the dark contours mark the $\widehat{L}_g=0.75$ hotspot cores used to calculate $\eta_{0.75}$.}
\label{fig:fracmaps}
\end{figure}

The aligned local FRAC-deficit fields are shown in Fig.~\ref{fig:fracmaps}, and their normalized high-deficit areas are reported in Table~\ref{tab:summary} and Fig.~\ref{fig:metrics}(c). RG3 exhibits the clearest compact joint-centered field. Its $\widehat{L}_g\ge0.75$ core occupies 1.49\% of the scanned area and remains concentrated around one dominant region after the four loose-bolt maps are aligned and averaged. RG1 and RG2 also have small core fractions, 2.87\% and 2.01\%, respectively, but their maps show weaker or less distinct joint-centered concentration than RG3. Thus, the similarly low global dissimilarities of RG1$\--$RG3 do not imply equivalent local behavior.

RG4$\--$RG6 show a different spatial pattern. Their normalized high-deficit areas increase to 9.63\%, 9.91\%, and 7.98\%, respectively, and the dark regions extend over several parts of the plate rather than remaining confined to one compact joint-centered region. These groups therefore combine larger global distortion with broader FRF redistribution. RG7 provides a contrasting case: despite a mean $1-\mathrm{CMAC}$ of 0.608, its thresholded core occupies only 0.81\% of the scanned area. The RG7 map nevertheless contains spatially distributed deficit features outside that small core. Consequently, the thresholded area quantifies the extent of the highest normalized deficits, but it does not by itself describe their absolute magnitude, position, or the remainder of the response field. The global coefficients, the normalized area, and the full local map should therefore be interpreted together.

\section{Discussion}
The global results show that retaining phase does not provide a uniform correction across the seven resonance groups; instead, it changes the separation between the low- and high-distortion regimes. For RG1$\--$RG3, the mean amplitude dissimilarity is 0.0507 and the mean complex dissimilarity is 0.0690, an absolute increase of only 0.0183 when phase is retained. For RG5$\--$RG7, the corresponding means are 0.2997 and 0.6103, an increase of 0.3107, which is approximately 17 times the increase observed for RG1$\--$RG3. At the individual-group level, $1-\mathrm{CMAC}$ is 2.26, 1.79, and 2.11 times $1-\mathrm{MAC}_a$ for RG5, RG6, and RG7, respectively, whereas the ratio is only 1.27 for RG4. The complex-MAC formulation is specifically intended to retain phase through complex response vectors \cite{Vacher2010}. Frequency-domain correlation of complex FRFs likewise preserves information beyond an amplitude-only comparison \cite{Perez2019}, and this type of correlation has been applied to damage quantification using measured FRFs \cite{Perez2021Ice}. In the present experiment, phase retention is therefore the principal discriminator between the intermediate RG4 response and the strongly redistributed RG5$\--$RG7 responses.

The uncertainty intervals and tracked frequencies show that this high-frequency sensitivity is also strongly case dependent. The mean width of the 95\% $1-\mathrm{CMAC}$ interval is 0.032 for RG1$\--$RG3 but 0.642 for RG5$\--$RG7, nearly a 20-fold increase; because only four bolt locations are resampled, this difference should be read as bolt-location variability rather than population-level uncertainty. The tracking table provides a physical reason for part of this variability: RG4 and RG5 select the same loose-case peaks at 7989~Hz for bolt 2 and 7985~Hz for bolt 4, while RG6 and RG7 share 8541~Hz for bolt 2 and 8564~Hz for bolt 3. These repeated assignments confirm that the high-frequency labels represent overlapping experimental response regions rather than uniquely identified modes. Peak picking alone cannot establish modal correspondence; modal identification requires the joint estimation and validation of frequencies, damping ratios, and mode shapes \cite{Reynders2012}. Recent automated modal-identification frameworks likewise separate physical modes from spurious numerical or measurement-related poles \cite{Chen2025AOMA}. Reporting $\Omega_a$ remains useful because dimensionless frequency scaling supports comparison among rectangular plates with different dimensions and bending stiffness \cite{Senjanovic2015}, and normalized-frequency formulations remain standard in modern thin- and thick-plate vibration analysis \cite{Eftekhari2021}. The coefficient nevertheless supplies frequency scaling, not mode identification.

The combined global and local results suggest that the resonance groups should be interpreted on two axes: the magnitude of whole-field FRF redistribution and the spatial concentration of the strongest local mismatch. RG1$\--$RG3 have a mean $1-\mathrm{CMAC}$ of 0.069 and a mean hotspot area of 2.12\%, defining a low-global-distortion regime. RG4$\--$RG6 have a mean hotspot area of 9.17\%, 4.32 times the RG1$\--$RG3 mean, and their maps show broad multi-region redistribution. The contrast between RG3 and RG5 is especially strong: RG5 has 8.55 times the complex dissimilarity and 6.65 times the normalized hotspot area of RG3. RG7 breaks any simple monotonic relation between the two descriptors: its $1-\mathrm{CMAC}$ value of 0.608 is 8.33 times that of RG3, yet its 0.81\% hotspot core is 46\% smaller. Conversely, RG4 retains 97\% of the RG5 hotspot area while reaching only 30\% of the RG5 complex dissimilarity. These comparisons show that global distortion and hotspot extent are complementary rather than interchangeable screening variables. FRF-shape localization methods similarly use the spatial organization of response mismatch as a diagnostic quantity \cite{Liu2009}, while complex frequency-domain correlation provides a separate measure of global response change \cite{Perez2019}. Dense LDV sampling is valuable because it directly reveals whether the mismatch is compact or distributed \cite{DiMaio2021}; full-field plate measurements have previously demonstrated the value of spatial vibration fields for damage identification \cite{Chen2018PlateLDV}. Recent FRF-based localization studies infer damaged regions through an inverse finite-element model \cite{Saito2023FRF} or through trained classifiers \cite{Ruiz2024FRFML}. The present workflow avoids both requirements and gives a directly interpretable screening map, but it does not estimate physical damage size, preload loss, or remaining joint capacity.

The normalized hotspot-area fraction is therefore a supporting spatial descriptor, not an independent damage index. Because each group is min$\--$max normalized before thresholding, $\eta_{0.75}$ measures the relative area of that group's strongest deficit core and discards the absolute FRAC-deficit scale. RG7 has the second-highest complex dissimilarity but the smallest hotspot area, whereas RG4 and RG5 have almost identical hotspot areas despite markedly different global dissimilarities. A robust interpretation should therefore use the pair $(1-\mathrm{CMAC},\eta_{0.75})$ together with the complete local map: the first quantity ranks whole-field redistribution, the second describes the relative spatial extent of the strongest core, and the map identifies its location and fragmentation.

The main limitations are the single plate, four \SI{0}{\newton\metre} cases, approximate free-free support, overlapping high-frequency tracking windows, and symmetry averaging with an excitation point that is not reflected together with the response fields. The bootstrap intervals consequently describe case-to-case variation only, and the fixed 0.75 threshold remains analyst selected. Future work should test additional geometries, preload levels, bolt layouts, and installed boundary conditions; evaluate tracking-window and threshold sensitivity; retain case-specific maps; and compare the RG assignments with independently identified modes or validated numerical mode shapes.

\section{Conclusion}
A focused full-field LDV screening procedure was applied to a four-bolt plate using the all-tight reference and four single-bolt \SI{0}{\newton\metre} cases. Seven trackable experimental resonance groups were selected from the spatial RMS spectrum and reported with one plate-normalized frequency coefficient. Global correlation showed a low-distortion set at RG1$\--$RG3, an intermediate response at RG4, and strong phase-aware distortion at RG5$\--$RG7. Local FRAC maps provided the spatial interpretation: RG3 formed the clearest compact joint-centered hotspot, while RG4$\--$RG6 produced substantially broader high-deficit regions. The normalized hotspot-area fraction complements, but does not replace, the global dissimilarity measures. The resulting workflow links resonance selection, global amplitude and phase correlation, and local FRF mapping in a form that can be reapplied without a trained classifier or specimen-specific node labels.

\section*{Declaration of competing interest}
The authors declare that they have no known competing financial interests or personal relationships that could have appeared to influence the work reported in this paper.

\section*{Data availability}
The experimental database used in this study has been deposited in Zenodo and is publicly available \cite{dataset_zenodo}. A companion data-descriptor manuscript \cite{datapaper} has been submitted for publication. The repository includes the measurement data, metadata, supporting documentation, and analysis scripts associated with the experiments.

\printcredits

\end{document}